\documentclass[10pt,letterpaper]{article}
\usepackage[top=0.85in,left=1.0in,footskip=0.75in]{geometry}

\usepackage[utf8]{inputenc}

\usepackage{cite}

\usepackage[right]{lineno}

\usepackage{microtype}
\DisableLigatures[f]{encoding = *, family = * }

\raggedright
\usepackage{changepage}

\usepackage[aboveskip=1pt,labelfont=bf,labelsep=period,singlelinecheck=off]{caption}

\makeatletter
\renewcommand{\@biblabel}[1]{\quad#1.}
\makeatother

\usepackage{lastpage,fancyhdr,graphicx}
\usepackage{epstopdf}
\usepackage{color}

\definecolor{Gray}{gray}{.25}

\usepackage{graphicx}

\usepackage{sidecap}

\usepackage{wrapfig}
\usepackage[pscoord]{eso-pic}
\usepackage[fulladjust]{marginnote}
\reversemarginpar

\begin{document}
\vspace*{0.35in}

\begin{flushleft}
{\Large
\textbf\newline{Vortical Flow Development in Round Ducts Across Scales for Engine Inlet Applications}
}
\newline
\\
Tamara Guimar\~{a}es\textsuperscript{1,*},
K. Todd Lowe\textsuperscript{2},
Walter F. O'Brien\textsuperscript{1}
\\
\bigskip
\bf{1} Turbomachinery and Propulsion Research Laboratory, Department of Mechanical Engineering, Virginia Tech, Blacksburg, VA 24061
\\
\bf{2} Kevin T. Crofton Department of Aerospace and Ocean Engineering, Virginia Tech, Blacksburg, VA 24061
\\
\bigskip
* tgbucalo@vt.edu

\end{flushleft}

\section*{Abstract}
Turbofan engine performance depends highly on the characteristics and conditions of the inlet flow. Swirl distortions, caused by non-uniformities in flows arising from boundary layer or ground/fuselage vortex ingestion are of concern and need to be fully understood to guarantee efficiency and safety of propulsion systems. To investigate a fundamental single-vortex distortion development in a duct at different Reynolds numbers, a StreamVane distortion generating device was designed and experimentally analyzed in a small-scale low-speed wind tunnel ($\mathrm{Re_{D}}~500\times10^3$) and in a full-scale engine testing rig ($\mathrm{Re_{D}}~3\times10^6$). Stereoscopic particle image velocimetry was used to measure the three-component velocity fields at discrete measurement planes downstream of the distortion device. Results show that the secondary flow is generated and develops very similarly in both scales, and is mostly driven by two-dimensional (2D) vortex dynamics. Induced velocities arising from the proximity of the vortex to the duct wall causes the vortex center to convect circumferentially around the duct, in the same sense as the vortex rotation, as it travels downstream. Small-scale turbulence results show small-scale instabilities related to the development of the vortex. This work shows that the development of this vane-generated, vortex-dominated flow is largely Reynolds number independent for the covered range, so that details of similar duct-bounded flows can be analyzed in depth in small-scale experiments, decreasing development efforts and cost.


\section*{Introduction}
Vortex-dominated flows are encountered in a number of applications across physical scales, and hold fundamental interest for a number of reasons. For instance, vortices generated from airplane wing tips, by bird wings, and by the movement of fish and other animals in the ocean, all influence how the fluids around those bodies behave locally. In a larger scale, vortices in the atmosphere are responsible for driving phenomena such as hurricanes and tornados. The work of Rockwell \cite{Rockwell1998} has confirmed the usefulness of Helmholtz's vortex theorems in practical flows such as the formation of leading-edge vortices on delta-wings and on the interaction of vortices with strain and surfaces. To note in Rockwell's and many others' work is the scale independence of vortex-dominated flows. The results from low Reynolds number experiments have been shown to match the aerodynamics encountered for full-scale vehicles, since the phenomena can be scaled purely on geometry and circulation of the vortex. In the present work, these same principles are applied to vortices confined inside aircraft engine inlet ducts in order to provide insight on leading-order flow scaling and effects of vortical flow ingestion by turbine engines.\

The performance of turbofan engines depends on the characteristics of the flow entering the engine. While the effects of total pressure and total temperature distortions in engine inlets have been studied and well-characterized for decades \cite{SAE1978, SAE1983, Schneck2013, Seddon1999}, swirl distortions are still not fully understood by the turbomachinery community. Swirl distortions are angular non-uniformities in the flow, which can be caused by several phenomena, such as boundary layer ingestion, ingestion of ground/fuselage vortices, the presence of cross wind, and adverse pressure gradients \cite{SAE2017}.\

Swirl distortions may affect the performance of the engine by modifying the incidence angle of the flow in the blades of various stages of the engine \cite{Fredrick2011}, and generating a temperature distortion behind the compressor due to work variations \cite{Genssler1987, Pazur1991, Sheoran2009}. Several investigations of the effects of swirl distortion have been performed, both computationally \cite{Anderson1993,Sheoran2009,Sheoran2012}, and experimentally. These experimental efforts have employed pressure probes in ducts \cite{Frohnapfel2015,Hoopes2013} or cascades \cite{Hobson1993}, and, more recently, particle image velocimetry \cite{MacManus2017,Murphy2010,Murphy2011,Zachos2016}.\

Single vortex distortions, in the form of tightly wound vortices, may arise in static conditions, due to the ingestion of a ground or a fuselage vortex, or due to cross winds \cite{DeSiervi1982}. This can cause debris to be ingested into the engine, increase blade erosion due to dirt ingestion \cite{Younghans1978}, and, since the core of the vortex is a region of total pressure deficit, this may cause engine surge \cite{Motycka1976}. Most of the investigations of vortex ingestion have focused on the mechanisms of formation of inlet vortices, or the generation of vortices inside the turbomachinery system \cite{AGARD1990,Sonoda1987}. Previous work on ground vortex ingestion by Brix et al. \cite{Brix2000}, and Murphy and MacManus \cite{Murphy2011} have provided important insight and conclusions about the generation of inlet vortices, and the relationship between ground clearance and relative speed for inlet vortex formation. However, no work has been done on the development of those vortices after their formation. To characterize the role of inlet vortex ingestions, it is necessary to understand the physics that drive the development of a fundamental vortical structure before it approaches the turbomachine components.\

In this work, a single vortex distortion was generated in two different scales to assess the dependence of the vortical flow characteristics and development on Reynolds number, and to describe the physics that drive the development of the flow. Stereoscopic particle image velocimetry (PIV) was used to measure the three-component velocity fields at discrete planes downstream of the distortion in small-scale low speed wind tunnel, and in the inlet of a full-scale research engine, far enough upstream of the engine to avoid fan effects. The experimental methods used in this work are presented in the following section, and a detailed analysis of the generated flow is presented in the Results section to show that the single-vortex distortion development is driven by two-dimensional (2D) vortex dynamics, and that the mean characteristics of the secondary flow are Reynolds number independent.\

\section*{Experimental Methods}
\label{sec:1}
\subsection*{The LOvort StreamVane Distortion Generator}
\label{sec11}
\begin{figure}
\centering
  \includegraphics[width=0.9\textwidth]{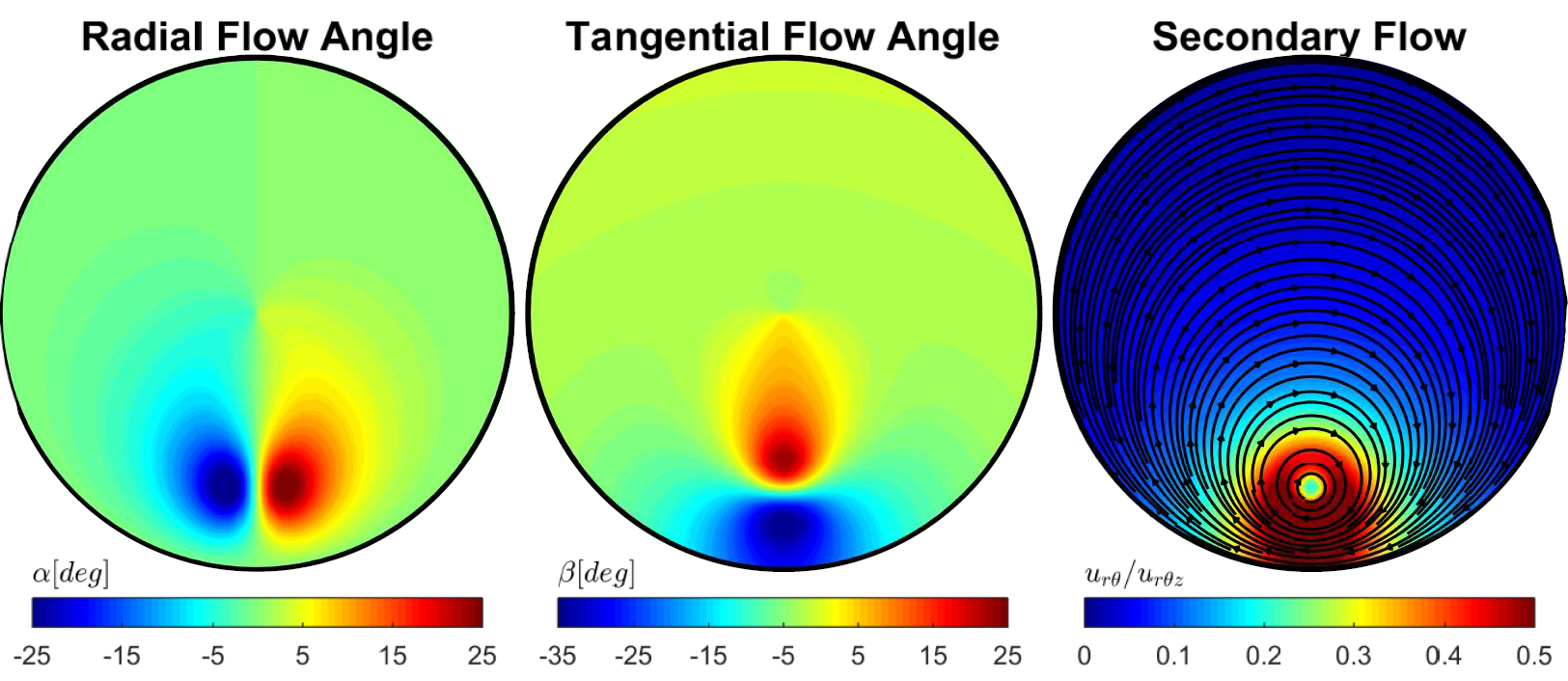}
\caption{Design flow angles and secondary flow at the exit plane (trailing edge) of the LOvort StreamVane.}
\label{Fig1}       
\end{figure}

A StreamVane\textsuperscript{TM} distortion  device was used to generate a single vortex distortion \cite{Guimaraes2016,Hoopes2013}, with the design inputs shown in Fig. \ref{Fig1}. The design flow is to be generated exactly downstream of the StreamVane, at 0 diameters downstream of its trailing edge.\

The strength of the vortex was chosen based upon simulation results of a highly distorted flow containing a tightly-wound vortex ingested by an engine on a hybrid-wing body aircraft concept \cite{Gatlin2012}, as described by Guimar\~{a}es et al. \cite{Guimaraes2017a}, and the vortex was modeled as a Lamb-Oseen vortex (therefore named LOvort), with the complex velocity described in Eq. \ref{eq:1}:

\begin{equation} \label{eq:1}
W(x+iy)=u_{x}-iu_{y}= \frac{\Gamma}{2i\pi}\frac{1}{x+iy}\left[1-exp\left(-\frac{|x+iy|^{2}}{a^{2}_{0}}\right)\right]
\end{equation}

\noindent where normalized circulation $\Gamma/D^{2}=-55.5 s^{-1}$ and vortex radius $a_{0}/D=0.05$. The location of the center of the vortex was chosen as 70\% of the duct radius, since Brix et al. \cite{Brix2000} determined that a vortex closer to the wall would be distorted by wall interactions. The wall boundary conditions were simulated by a complementary image vortex (with opposite circulation) by adding another term to the complex velocity field, as shown in Eq. \ref{eq:2}. The throughflow velocity was designed as constant.

\begin{equation} \label{eq:2}
W(x+iy)= u_{x}-iu_{y}= \frac{\Gamma}{2i\pi}\frac{1}{x+iy}\left[1-exp\left(-\frac{|x+iy|^{2}}{a^{2}_{0}}\right)\right]+ \frac{-\Gamma}{2i\pi}\frac{1}{x-iy}\left[1-exp\left(-\frac{|x-iy|^{2}}{a^{2}_{0}}\right)\right]
\end{equation}

\begin{figure}
\centering
  \includegraphics[width=0.6\textwidth]{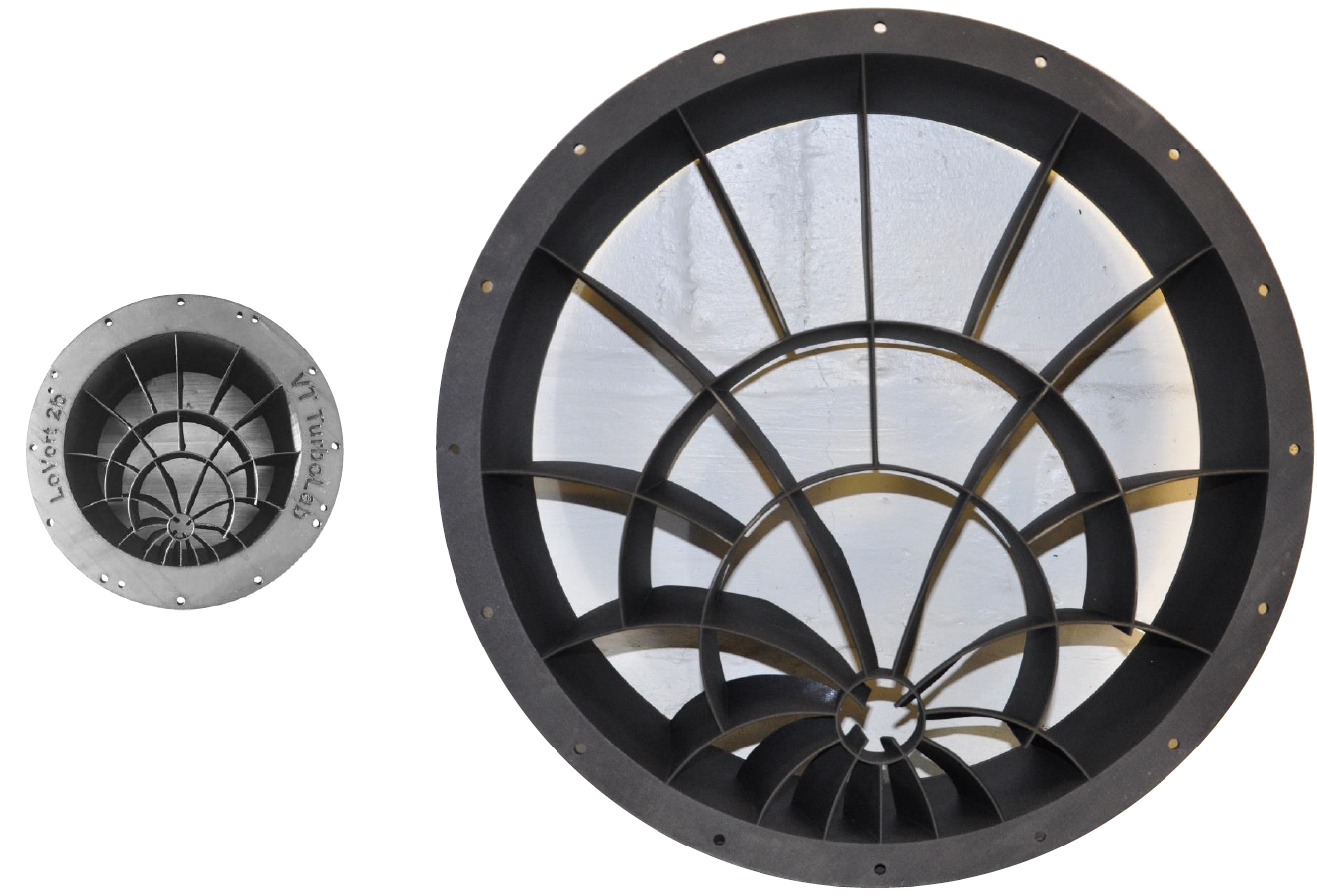}
\caption{Small-scale StreamVane (left), and full-scale StreamVane (right) presented to scale.}
\label{fig:2}
\end{figure}

All plotted results will be presented from a forward-looking-aft (FLA) point of view, with the tangential velocity component being positive in the counterclockwise direction, and the radial velocity component being positive radially outward. Radial flow angles ranged from -25 to 25 degrees, and tangential flow angles from -35 to 25 degrees. The resulting small-scale StreamVane distortion device is shown in Fig. \ref{fig:2} (left). Its diameter was 0.15 m (6 inches), maximum chord length was 0.082 m (3.25 inches) and it was built in Acrylonitrile Butadiene Styrene (ABS) through additive manufacturing.\

Once the results from the small-scale experiment were verified and satisfactory, the StreamVane was scaled up for experiments in the full-scale rig, and built using additive manufacturing in black ULTEM 9085. The full-scale StreamVane is also shown in Fig. \ref{fig:2} (right). The diameter of the full-scale StreamVane was 0.534 m (21 inches), and its maximum chord length was set to 0.18 m (7.2 inches). The overall design was scaled directly from the small-scale, yielding vanes longer than 0.18 m, which were manually cropped in the CAD design before manufacturing, since the longer vanes in regions of low levels of turning generated thicker wakes. The thickness of the vanes was set to 3.175 mm (1/8 inch) to minimize the effects of the wakes generated by the vanes in the flow.\

\subsection*{Small-Scale Wind Tunnel}
\label{sec12}
\begin{figure}
\centering
  \includegraphics[width=1.0\textwidth]{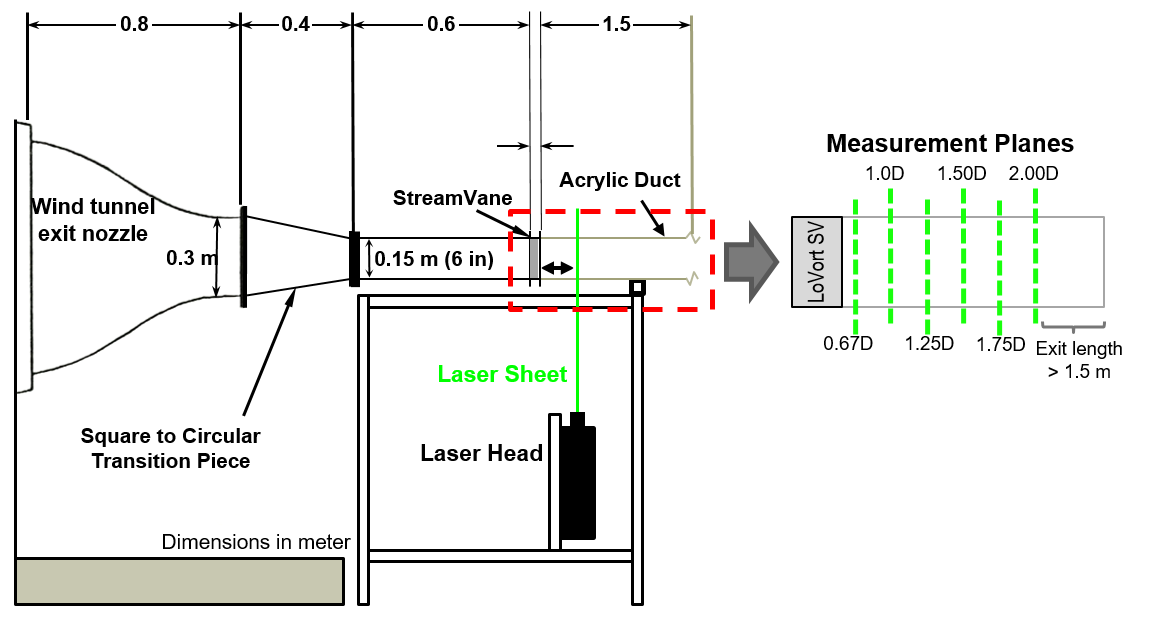}
\caption{Small-scale wind tunnel setup and measurement planes.}
\label{fig:3}
\end{figure}

The small-scale experiment was performed in the Virginia Tech Turbomachinery and Propulsion Research Laboratory wind tunnel, which is equipped with a square-to-round outlet designed to fit a 0.15 m diameter pipe. The schematics of the experimental setup, including the tunnel outlet, are shown in Fig. \ref{fig:3}. The upstream bulk velocity for this tunnel was $49.3 \pm 0.7~\mathrm{m/s}$ (Mach 0.14), and duct Reynolds number of around 500,000, where $Re_{D}=u_{\infty}D/\nu$, and $\nu$ for 300 K is $1.57\times10^{-5}~ \mathrm{m^{2}/s}$.

\begin{figure}
\centering
  \includegraphics[width=0.9\textwidth]{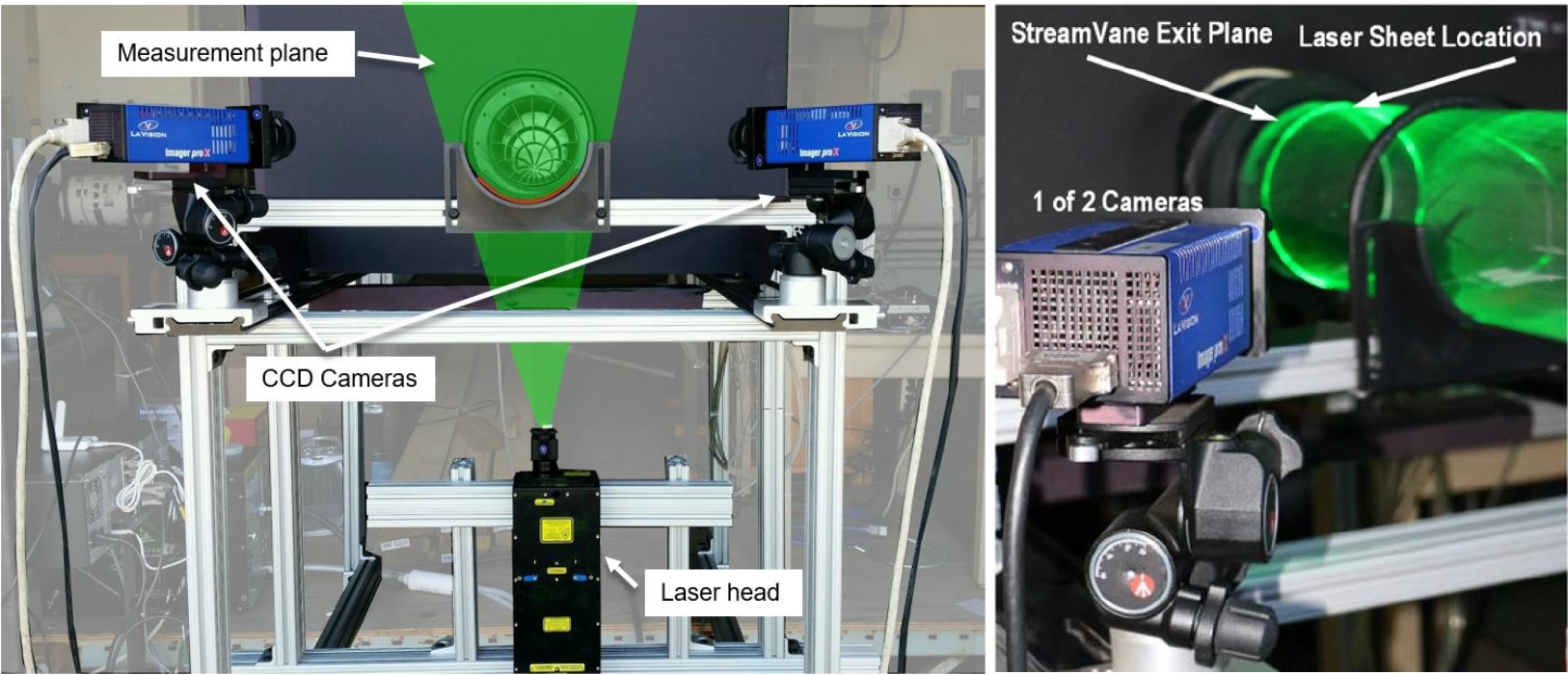}
\caption{PIV experimental setup (left), and details of the laser sheet intersecting with the acrylic pipe (right \cite{Sanders2016}).}
\label{fig:4}
\end{figure}

Stereoscopic particle image velocimetry was used to measure the three-component velocity fields at six discrete measurement planes downstream of the StreamVane: 0.67, 1.00, 1.25, 1.50, 1.75, and 2.00 diameters downstream of the leading edge (StreamVane thickness was 0.54D). Two LaVision Imager Pro X 4M cameras were placed in a stereo configuration for imaging the flow. They were equipped with 50 mm Nikon lenses and Scheimpflug adapters, yielding a viewing image size of approximately 0.16 x 0.16 $\mathrm{m^{2}}$ to ensure the entire inside of the duct was captured. A thousand images were acquired in each plane at 4 Hz, with a time delay of 18 $\mu s$ between the first and second frame of the image. The laser used to illuminate the particles was a Quantel Evergreen 200 double-pulsed Nd:YAG laser, rated 200 mJ/pulse, emitting at 532 nm. A LaVision collimator and a -10 mm cylindrical lens were placed at the exit of the laser to treat the laser beam into a 3 mm (0.12 in) thick laser sheet for the measurements. The flow was seeded from upstream of the wind tunnel blower with 0.2-0.3 $\mu\mathrm{m}$ diameter particles of fluid A (glycerol + water) using a Concept Engineering Ltd Colt 4 Seeder. The PIV measurement section of the wind tunnel consisted of a 1.52 m long optically clear cast acrylic pipe with 0.15 m in diameter, shown in Fig. \ref{fig:4}.\

\subsection*{Full-Scale Testing Rig}
\label{sec13}

The full-scale testing rig was comprised of tunnel sections and a bellmouth for inlet flow conditioning, one particle seeding distribution device, the StreamVane installed in a custom-built rotator, a customized test section with optical access for the cameras and the laser sheet, connection tunnel sections and the Pratt and Whitney Canada JT15D-1 research engine mounted to a test stand. Measurement planes were defined far enough upstream of the fan, so that fan effects were avoided. Axial average bulk velocity for this setup was 87 m/s (70\% corrected fan speed ($N/\sqrt{T}$), Mach 0.25), yielding a duct Reynolds number of 3 million, based on a duct diameter of 0.53 m (21 inches).\

\begin{figure}
\centering
  \includegraphics[width=0.9\textwidth]{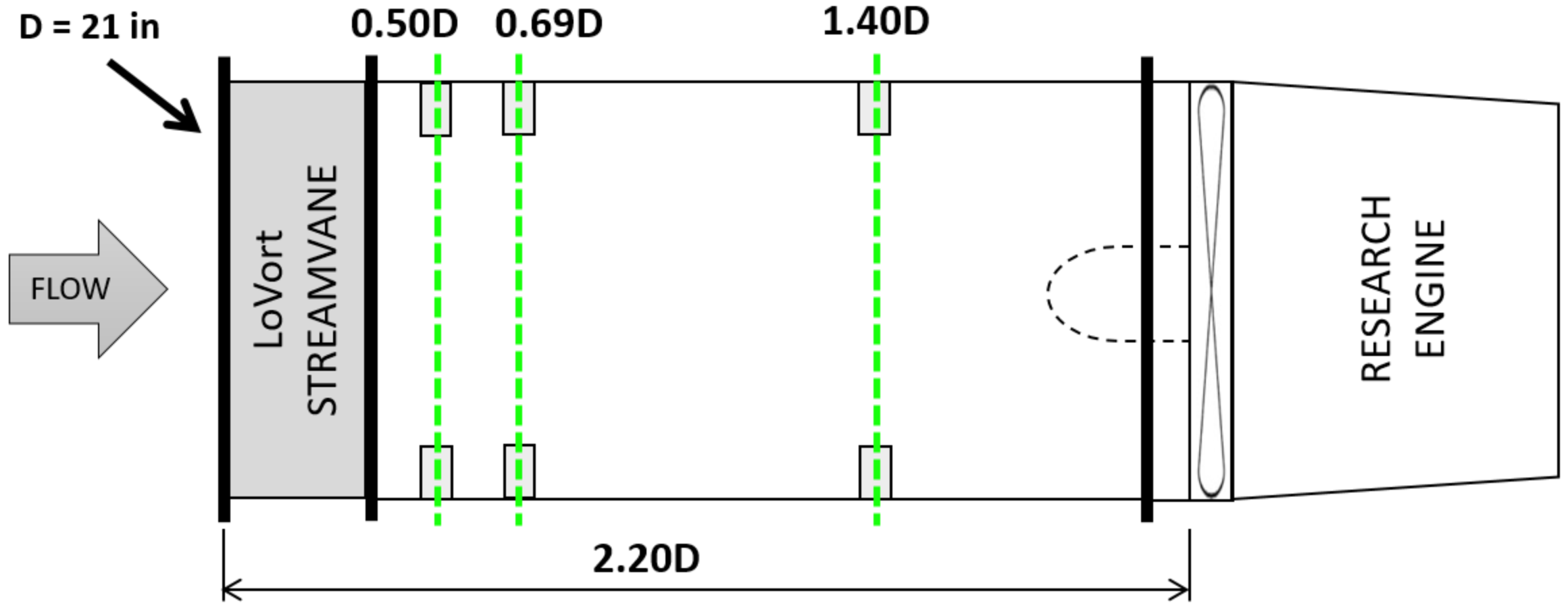}
\caption{Full-scale measurement planes relative to the leading edge of the StreamVane.}
\label{fig:5}
\end{figure}

Two LaVision Imager Pro X 4M CCD cameras (2048 x 2048 px, 7.4 $\mu \mathrm{m}$ pixel size) were mounted in a stereoscopic configuration, equipped with 24 mm Nikon lenses with aperture set to 5.6, and Scheimpflug adapters. Data were taken in three discrete measurement planes downstream of the StreamVane: 0.50, 0.70, and 1.40 diameters relative to the StreamVane leading edge (StreamVane thickness was 0.34D). The position of the StreamVane relative to the fan face was maintained constant at 2.20 diameters. For the 0.50 and 0.70D measurement planes, the cameras were facing upstream of the flow, and for the 1.40D they were facing downstream. Due to the limited optical access available in the full-scale environment, data were taken in slices. The field of view for each slice was maximized to ensure enough overlap, and the StreamVane was rotated by 30 degrees between each measurement. The same laser system was used as the small-scale experiment. A thousand images were acquired in each measurement slice at 4 Hz, with a time delay of 5 $\mu \mathrm{s}$ between the first and second frame of the image. Further details of the setup were previously presented in Guimar\~{a}es et al. \cite{Guimaraes2018a}.\

Table \ref{tab:1} summarizes the main properties of the small-scale and the full-scale experiments.\

\begin{table*}
\centering
\caption{Summary of Small- and Full-Scale Experiment properties}
\label{tab:1}
\begin{tabular}{cccccc}
\hline\noalign{\smallskip}
Experiment & Duct & Average Bulk & Reynolds & Mach & Measurement Planes  \\
	       & Diameter & Axial Velocity & Number & Number \\
\noalign{\smallskip}\hline\noalign{\smallskip}
Small-Scale & 0.15 m (6 in) & 50 m/s & 0.47 M & 0.14 & 0.67, 1.00, 1.25, 1.50, 1.75, 2.00D\\
Full-Scale & 0.53 m (21 in) & 87 m/s & 2.90 M & 0.25 & 0.50, 0.69, 1.40\\
\noalign{\smallskip}\hline
\end{tabular}
\end{table*}

\section*{Analysis}
\label{sec:2}

\subsection*{Small-Scale PIV Processing}
\label{sec21}
Images were first processed in the LaVision DaVis software using stereo cross-correlation for obtaining velocity vectors. A sliding background was subtracted, and particle intensity normalization was performed, at a 2-pixel scale. Then, multi-pass processing was used, where the first pass consisted of a 128 x 128 square window with 75\% overlap, and the following 2 passes consisted of a 64 x 64 square window with 75\% overlap. A built-in median filter was used to remove and replace spurious vectors based upon the neighboring vectors. Vectors that did not fall within an interval of the average $\pm$2 times the standard deviation were removed, and vectors with a peak ratio (the ratio of the largest correlation peak to the second largest correlation peak in an interrogation window) of less than 1.3 were deleted. One pixel was equivalent to 76 $\mu \mathrm{m}$ for this configuration, yielding a spatial resolution of 4.9 mm for the smaller interrogation window.\

Since the laser sheet was entering and exiting the acrylic pipe, laser glare was an issue. The data was then further processed to ensure that spurious and biased vectors were deleted. A statistical analysis of the flow was performed, and it was found that the most relevant parameters for data filtering in this experiment were data count, axial velocity variance ($\sigma_{z}^{2}$), horizontal velocity variance ($\sigma_{x}^{2}$), axial velocity magnitude ($u_{z}$), and horizontal velocity magnitude ($u_{x}$). Based on the cameras being placed along the horizontal axis of the experiment, it was expected that the $u_{x}$ and $u_{z}$ components of velocity would present more uncertainty and spurious data, since the slightest misalignment between calibration plane and laser sheet position would cause erroneous correlations. In-plane velocity components ($u_{x}$ and $u_{y}$) greater than 27 m/s were filtered, as were axial velocity components ($u_{z}$) smaller than 25 m/s. The velocity variations were limited to $\pm$40 m/s in the most turbulent case and $\pm$14 m/s in the case with less turbulence.

\subsection*{Full-Scale PIV Processing}
\label{sec22}
One of the main challenges with this experimental setup is seeding, as presented previously by Nelson \cite{Nelson2014}. For this particular distortion case, it was a challenge to consistently seed the core of the generated vortex. Because of that, it was not possible to obtain turbulence information about the flow with sufficient confidence to draw any conclusions, however the amount of data collected was enough to observe the mean characteristics of the flow.\

LaVision Davis was also used to process the images using stereo cross-correlation. Background images were taken in between the measurement slices, averaged, and then subtracted from the PIV images. Then, multi-pass processing was used, where the first pass consisted of a 128 x 128 square window with 50\% overlap, and the following 2 passes consisted of a 64 x 64 round window with 75\% overlap. A built-in median filter was used to remove and replace spurious vectors based upon the neighboring vectors. Vectors that did not fall within an interval of the average $\pm$2 times the standard deviation were removed, and vectors with a peak ratio of less than 1.3 were deleted. Vectors in which the in-plane velocities did not fall within $\pm$30 m/s, or the axial velocity was smaller than 25 m/s or greater than 125 m/s, were deleted. One pixel was equivalent to 150 $\mu \mathrm{m}$ for this configuration, yielding a spatial resolution of 9.5 mm for the smaller interrogation window.\

To obtain full velocity profiles in each measurement plane, the slices of data were stitched together in MATLAB through the polar binning method described by Guimar\~{a}es et al. \cite{Guimaraes2018a}. 

\subsection*{Uncertainties}
\label{sec23}
Quantifying PIV uncertainty is always a challenge, due to the amount of variables that may contribute to increasing the experimental uncertainty of the setup, which are complex and not easily taken into consideration. According to Raffel et al. \cite{Raffel2007}, PIV data under ideal conditions can have a resolution of up to about 0.1 pixels, limited by the typical correlation peak interpolation. With that, an uncertainty analysis based on the uncertainty of the measurement in the particle displacement can be performed. Based on the scale factors calculated by the software, the displacement uncertainty is 8 $\mu$m for the small-scale experiment, and 0.015 mm for the full-scale. Assuming no uncertainty in the time measurements, the uncertainty in velocity would be 0.42 m/s, 0.9\% of the small-scale bulk velocity, and 3.0 m/s for full-scale, 3.5\% of the bulk velocity. For flow angle calculations, the uncertainty can be obtained from their definitions ($\alpha=tan^{-1}u_{r}/u_{z}$ and $\beta=tan^{-1}u_{\theta}/u_{z}$), resulting in an uncertainty $\sqrt{2}$ times higher than the velocities, therefore $\pm1.2$\% for the small-scale, and $\pm4.9$\% for full-scale, using a small angle approximation. These results are consistent with the uncertainty values calculated by the LaVision software \cite{Wieneke2015}, which were mostly lower than 0.5 m/s for the small-scale, and 3 m/s for the full-scale experiment.

\section*{Results and Discussion}
The LOvort StreamVane consists of a single vortex distortion, centered at about 70\% of the fan blade radius, and located at the bottom dead center of the duct, similar to what might be found in the inlet of a turbofan engine ingesting a ground vortex, or subjected to cross winds in embedded configurations. Two distinct experiments were performed to analyze the behavior of the generated vortex: a small-scale low speed wind tunnel experiment, and a full-scale test. The results to be presented indicate that the mean flow development is Reynolds number independent for the covered range, and that secondary flow development is driven by 2D vortex dynamics.\

In the mean-flow results discussed following, the design targets at the StreamVane exit for secondary velocities and flow angles are presented along with small- and full-scale experimental results. Small-scale experimental results are presented at planes $z =$ 0.67D and 1.50D, while full-scale results are presented at the closest planes for comparison, $z=$ 0.50D, 0.69D, and 1.40D. Flow measurements at 0.50D in full-scale are different from the design, since the flow has already started to develop. The black outer ring in the figures represents the duct inner diameter, and blank regions inside the ring are where data was filtered due to lack of quality caused by laser glare in the duct/windows.\

\begin{figure}
\centering
  \includegraphics[width=0.7\textwidth]{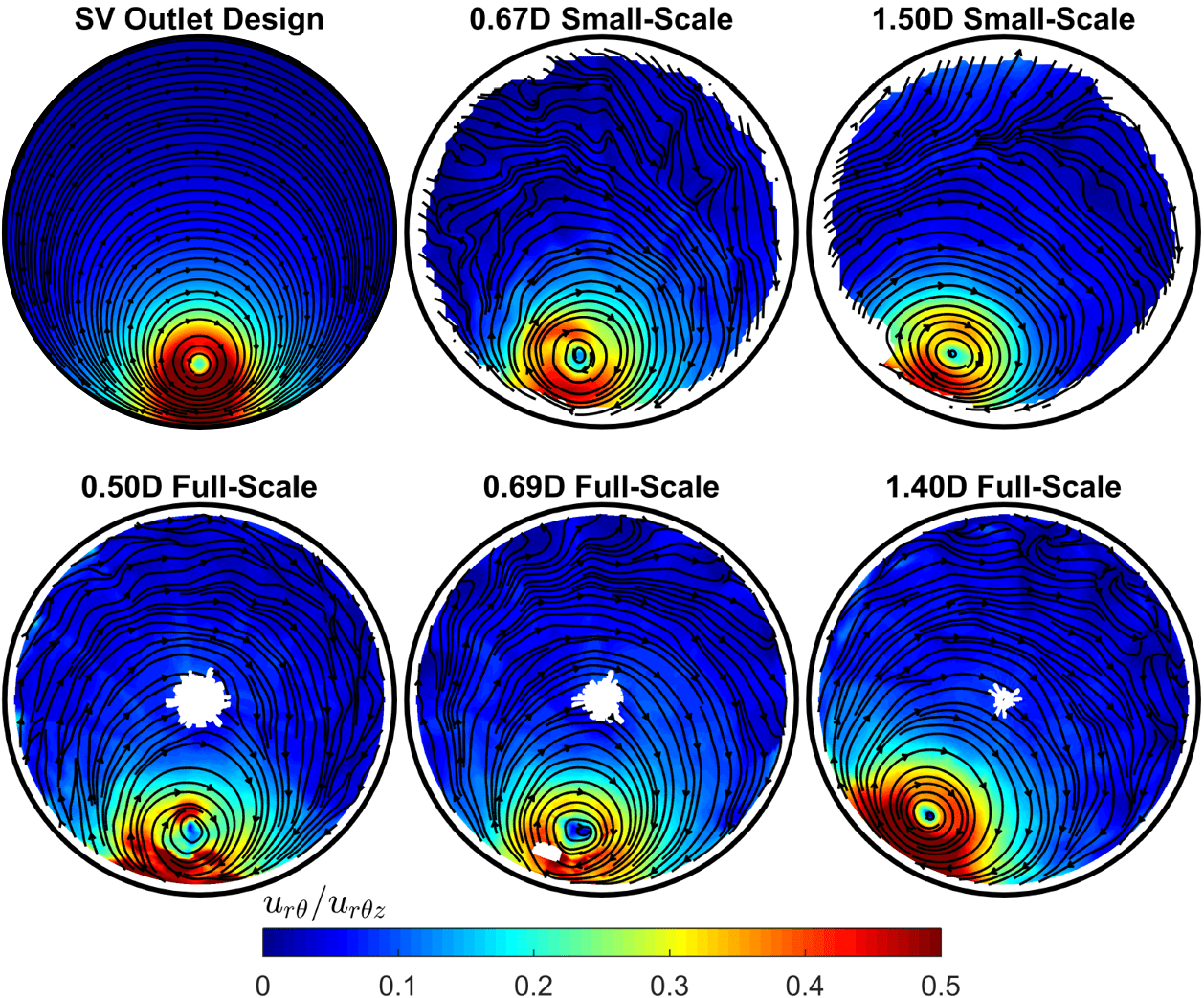}
\caption{Secondary velocity profiles. Lines represent the mean direction of the flow and the color plot shows the velocity intensities, normalized by the average bulk axial velocity.}
\label{fig:6}
\end{figure}

The mean normalized secondary flow velocity profiles shown in Fig. \ref{fig:6} were obtained by averaging the in-plane components of the velocity field ($u_{r\theta}\equiv \sqrt{u_{r}^{2}+u_{\theta}^{2}}$), and dividing them by the mean local total velocity ($u_{r\theta z}\equiv \sqrt{u_{r}^{2}+u_{\theta}^{2}+u_{z}^{2}}$). The StreamVane is manufactured to generate the design profile at the exit of the device, which would be 0D downstream of the trailing edge. Due to limitations in the experimental setup, it was not possible to take measurements at that location, so the first measured plane downstream was at 0.50 diameters downstream of the leading edge in the full-scale experiment. At that plane, it is possible to see that the flow has started convecting due to the interactions with the wall, and the vortex has migrated in the same direction as the vortex rotation by less than 5 degrees. It is also observed that there are three concentrated regions of higher in-plane velocity, inside the vortex, suggesting that the core of the vortex at that plane comprises three smaller vortices. This will be explained later in this section, when the axial velocity profiles are discussed.\

\begin{figure}
\centering
  \includegraphics[width=0.7\textwidth]{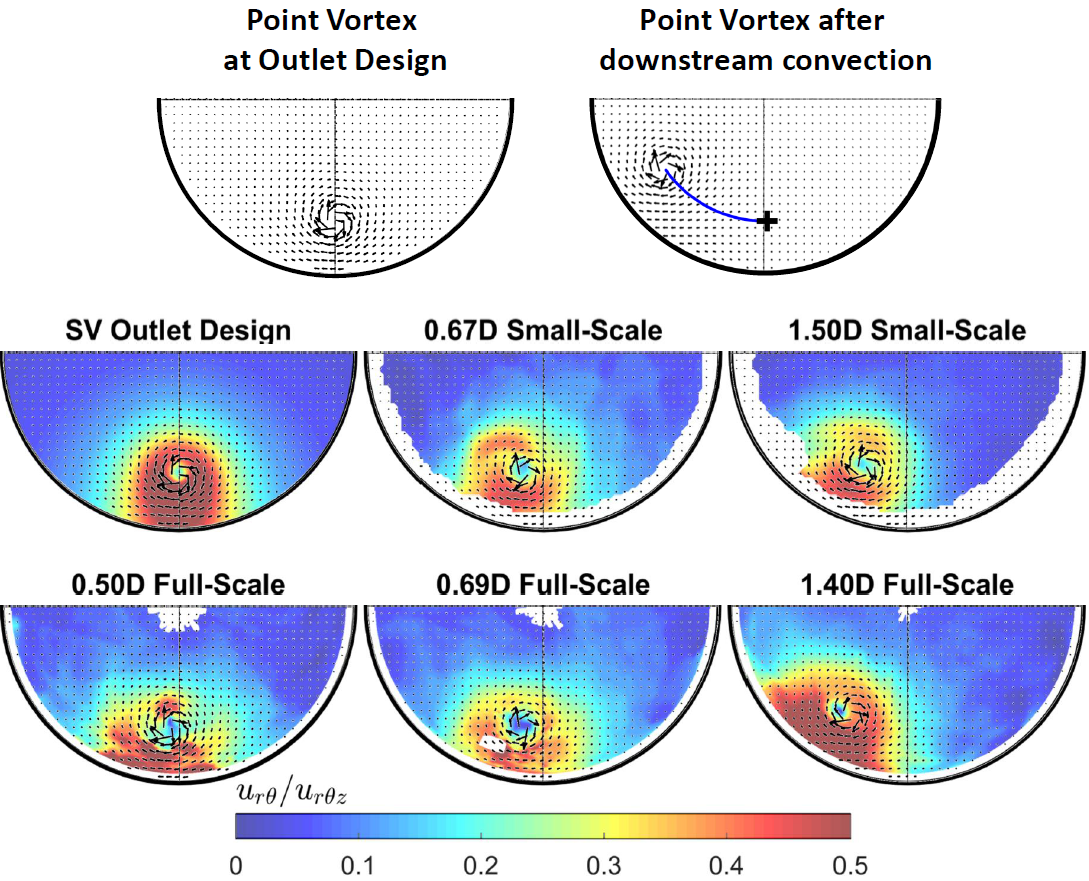}
\caption{Point vortex model downstream convection example. Vectors from simplified image vortex model, contours for secondary velocity magnitude from small-scale and full-scale experiments.}
\label{fig:7}
\end{figure}

\begin{figure}
\centering
  \includegraphics[width=1.0\textwidth]{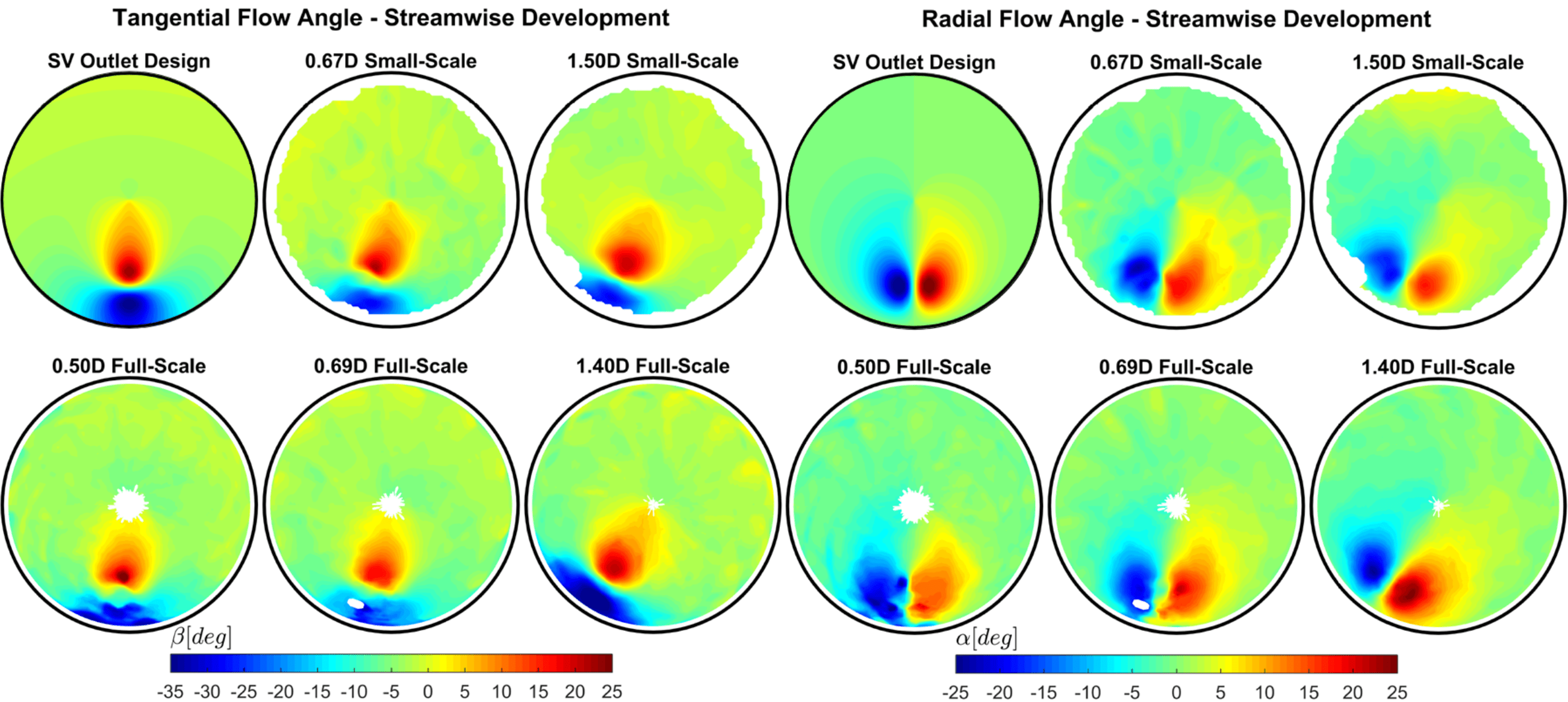}
\caption{Tangential (left), and radial (right) flow angle profiles.}
\label{fig:8}
\end{figure}

When focusing on the secondary velocity profiles, it seems that the vortex has maintained its overall strength and shape throughout the development, although the magnitude of the velocity around the vortex is not as uniform as design, with significantly higher values between the core and the wall. This indicates that the vortex development is highly governed by two-dimensional vortex dynamics, which state that for an incompressible flow, the fluid velocity can be determined by the vorticity of the flow, up to an irrotational far field component through the Biot-Savart law  \cite{Saffman1992}. It is also possible to observe some secondary flow motion from the vortex sheets being shed from the vanes of the StreamVane in the low turning regions of the 0.50D and 0.67D planes. They seem to mix quite quickly, and are almost unobservable at the 1.50D plane. The small-scale data have previously been successfully compared to an inviscid and incompressible low order flow development model, in which the in-plane velocity set as the initial condition was determined from the same goal profiles used to design the experimental distortion screen, and the axial development of the flow was treated as the time evolution of the profile \cite{Schneck2017}. The similarities of the small-scale flow with the full-scale experiment indicate that there are no further physics that need to be described for the higher Reynolds number flow that is far enough upstream of the engine fan face and outside of the duct inner-diameter boundary layer. This has also been previously observed by Sanders et al. \cite{Sanders2016} for a more complex StreamVane pattern, and compared to computational fluid dynamics results.\

To further test this hypothesis, a simple point vortex convection model was generated to compare to the experimental results. The point vortex was simulated in the same position as the vortex in the design profile, shown in Fig. \ref{fig:7} (top), and the circular boundary of the duct was generated by means of an image vortex placed outside of the duct diameter (not shown in figure) \cite{Buehler2002, Pozrikidis2011}. The Euler method was used to solve the ordinary differential equation $dx/dt=U(x,t)$ using the initial position of the vortex as the initial condition. As flow develops downstream, the vortex convects along the duct due to the induced velocity imposed to it by the image vortex that has the complex velocity 

\begin{equation} \label{eq:3}
W(x+iy)=\frac{-i\Gamma_{image}}{2\pi}\frac{1}{\left[\left(x+iy\right)_{real}-\left(x+iy\right)_{image}\right]}
\end{equation}

The simulated vortex position for each experimental measurement plane was calculated and plotted as vectors on top of the secondary velocity profiles presented previously in Fig. \ref{fig:6}, and most small- and full-scale experimental data seem to agree well with the model. The only plane that is slightly off is the further downstream full-scale plane, 1.40 D, indicating that the flow at that point in full-scale may be already subjected to some influence from the engine.\

The StreamVane is designed to generate swirl, which means it is optimized to meet tangential flow angle parameters. The resulting tangential and radial flow angles for this experiment are shown in Fig. \ref{fig:8}. With these details of the flow, it is possible to observe more of the characteristics of the vortex and its development in the duct. The tangential flow profiles show that the higher absolute values seem to all migrate with the vortex, reinforcing the fact that the overall shape of the vortex is preserved. The maximum and minimum flow angle values do not vary drastically, indicating little changes in the characteristics of the in-plane flow. The radial flow angles present a little more of the influence of the vanes in the generation of the vortical structure of the flow. These results reinforce the dominance of mean-flow vortex-dominated convection, with little influence from viscous or turbulence effects.\

\begin{figure}
\centering
\includegraphics[width=0.4\textwidth]{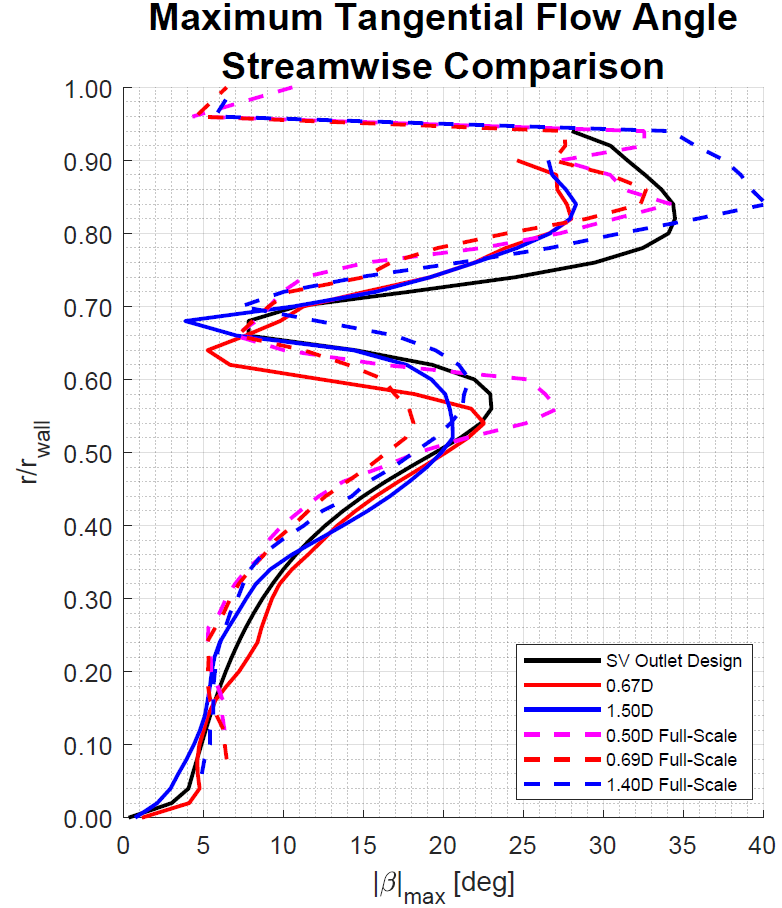}
\includegraphics[width=0.4\textwidth]{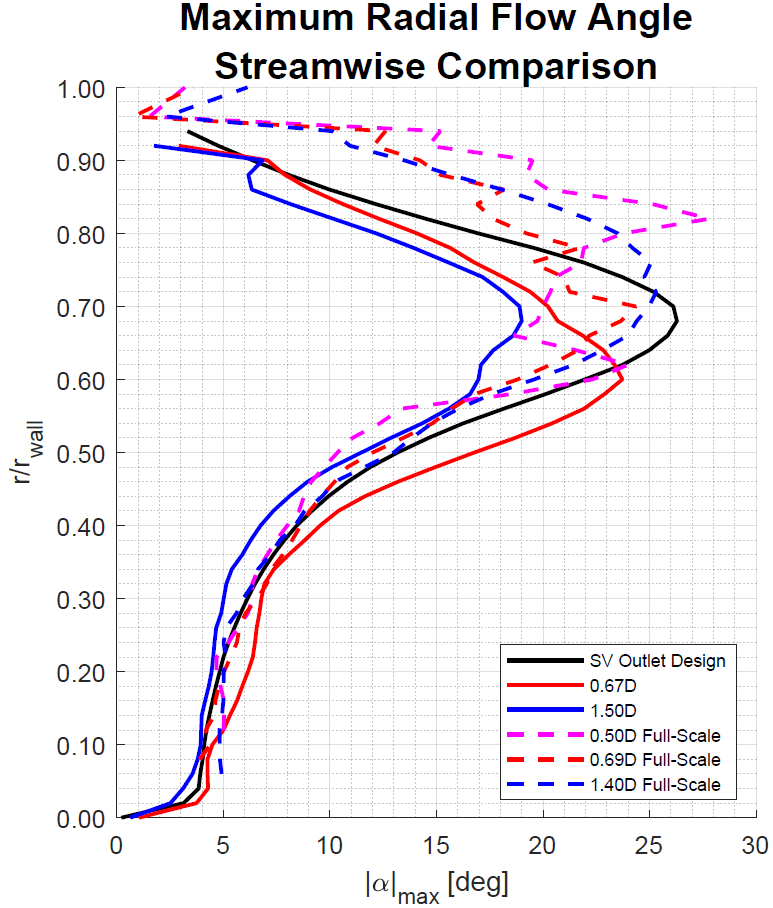}
\caption{Maximum absolute (a) tangential and (b) radial flow angle comparison between small and full-scale experiments.}
\label{fig:9}
\end{figure}

\begin{figure}
\centering
  \includegraphics[width=0.5\textwidth]{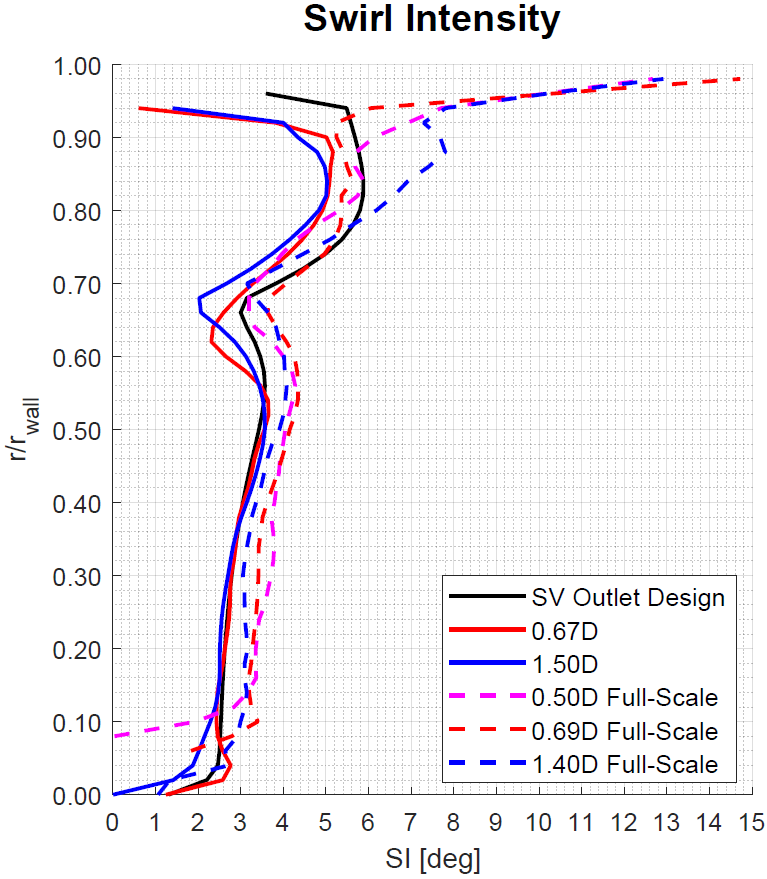}
\caption{Swirl intensity comparison between small and full-scale experiments.}
\label{fig:10}
\end{figure}

Figure \ref{fig:9} presents a comparison between the maximum absolute tangential and radial flow angles across the radius of all the measurement planes presented in Fig. \ref{fig:8}. The full-scale StreamVane seem to overshoot the higher angles, relative to the small-scale flow angles, but since the near-wall data was filtered for the small-scale experiment, this cannot be verified. The overall shape of the distortion distribution is maintained across scales, as shown previously for more complex geometries by Guimar\~{a}es et al. \cite{Guimaraes2018a} and Sanders et al. \cite{Sanders2016}. Most of the other regions of the radius seem to agree well across scales, mainly when taking into account the uncertainty of the measurements (plots for the same values with error bars are presented in the Appendix). It appears that the distortion is closer to the wall in the full-scale measurements, which may be responsible for part of the differences presented between the scales, mainly in the radial flow angle comparisons.\

Swirl intensity is one of the swirl descriptors defined by the SAE S-16 Committee to quantify inlet swirl distortions. It is defined as $SI_{i}=\frac{SS_{i}^{+}\times \theta_{i}^{+}+|SS_{i}^{-}| \times \theta_{i}^{-}}{360}$, where $SS_{i}^{+}$ is the integrated swirl angle over the $\theta_{i}^{+}$ extent, and it is used to capture the predominant swirl angle in degrees at the measurement plane \cite{SAE2017}. Calculated swirl intensities for the same measurement planes are shown in Fig. \ref{fig:10}. As with the maximum flow angles, for the majority of the radius of the measurement planes, the swirl intensity angles of the full-scale measurement planes are higher than the small-scale. We observe a general trend to overshoot the magnitude of the higher angles in full-scale, which is a feature mainly of the small differences in the printing of the StreamVane devices, and have no connection to the development of the flow. The main conclusion to be drawn from these flow angle measurements is that a comparable secondary flow profile is being generated in two different scales, and analyzing the development of the small-scale flow is enough to gain insight on how the flow will develop in the full-scale environment, before it interacts with the features of the turbofan engine, decreasing the costs of the experimental work involved.\

\begin{figure}
\centering
  \includegraphics[width=0.8\textwidth]{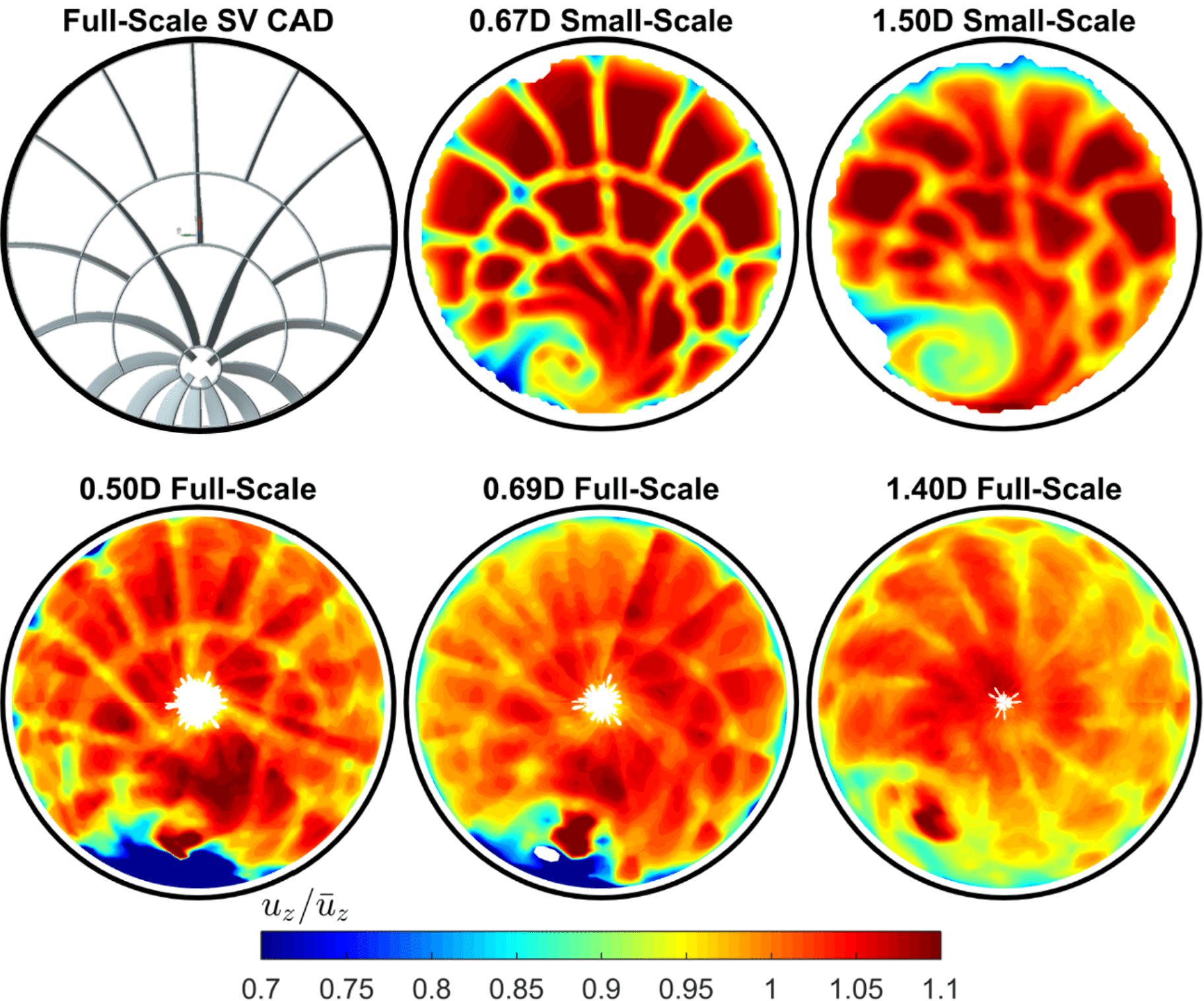}
\caption{Full-scale StreamVane model, small-, and full-scale axial velocity profiles normalized by the average bulk axial velocity.}
\label{fig:11}
\end{figure}

This StreamVane was designed with a constant axial velocity component. However, adding a distortion device comprised of vanes to generate turning will have an influence in the axial velocity component, mainly due to blockage. To better observe the effects of the vanes to the flow, and also to the distortion, the mean axial velocity profiles, normalized by the average bulk axial velocity, are presented in Fig. \ref{fig:11}, alongside with the CAD model of the StreamVane. The full-scale measurements for the axial velocity component were not as satisfactory as the in-plane components, and there is some spurious data left from the effects of glare and the stitching process, but the main overall features of the flow are maintained and can be observed, especially when taking the small-scale results as a reference. The 0.50D plane presents a high axial velocity deficit in the bottom dead center, caused by the high blockage introduced to the flow by the array of vanes that generate the bottom of the vortex. The wakes of the vanes are very well delimited in the 0.50D full-scale and the 0.67D small-scale measurement planes, and they start to mix as the flow develops downstream, which is expected for low Mach number flows. More interestingly, it is possible to observe the axial development of the vortex. This is not as clear when looking at the full-scale measurements, but at the 1.40D plane it is possible to observe some of that, while also observing a region of high axial velocity in the center of the vortex. This may be a result of the higher Reynolds number, presenting less effective blockage as the flow passes through the center of the vortex-generating ring in the StreamVane, and taking longer to mix, being a result of the distortion device, not of the flow development.\

\begin{figure}
\centering
  \includegraphics[width=0.9\textwidth]{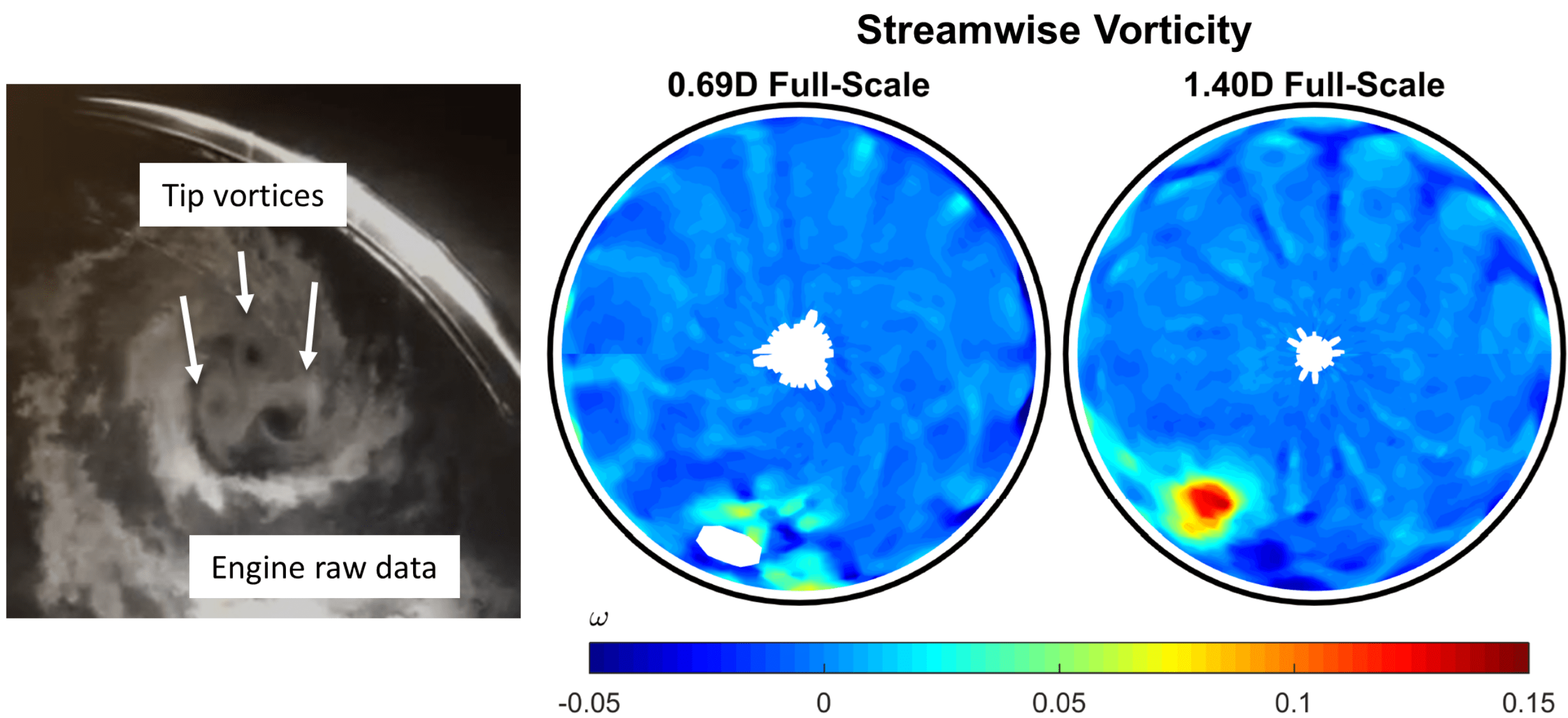}
\caption{Raw image showing three tip vortices generated by the StreamVane at the 0.50D full-scale measurement plane, and streamwise vorticity for full-scale data indicating a multi-core vortex at 0.69D.}
\label{fig:12}
\end{figure}

\begin{figure}
\centering
  \includegraphics[width=1.0\textwidth]{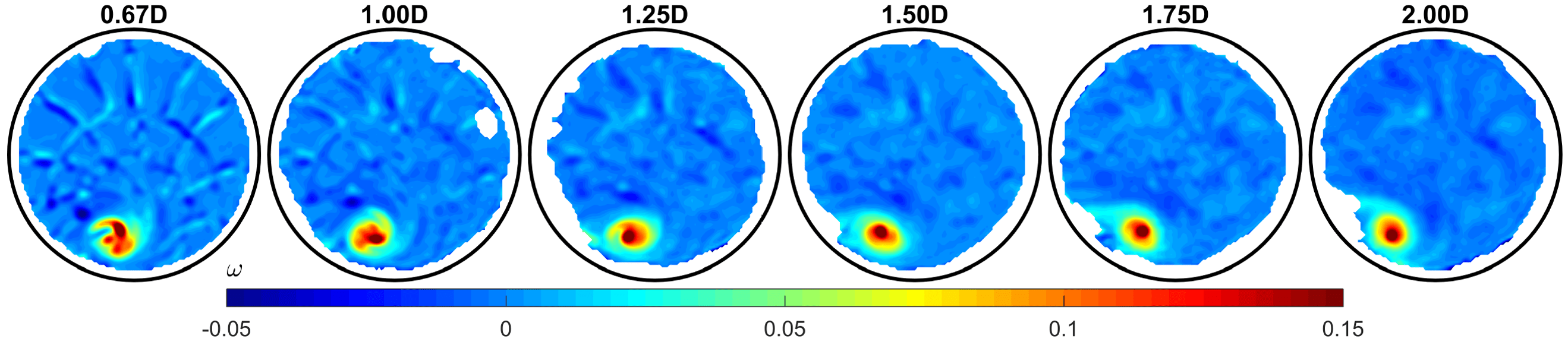}
\caption{Mean streamwise vorticity of the small-scale results.}
\label{fig:13}
\end{figure}

\begin{figure}
\centering
  \includegraphics[width=1.0\textwidth]{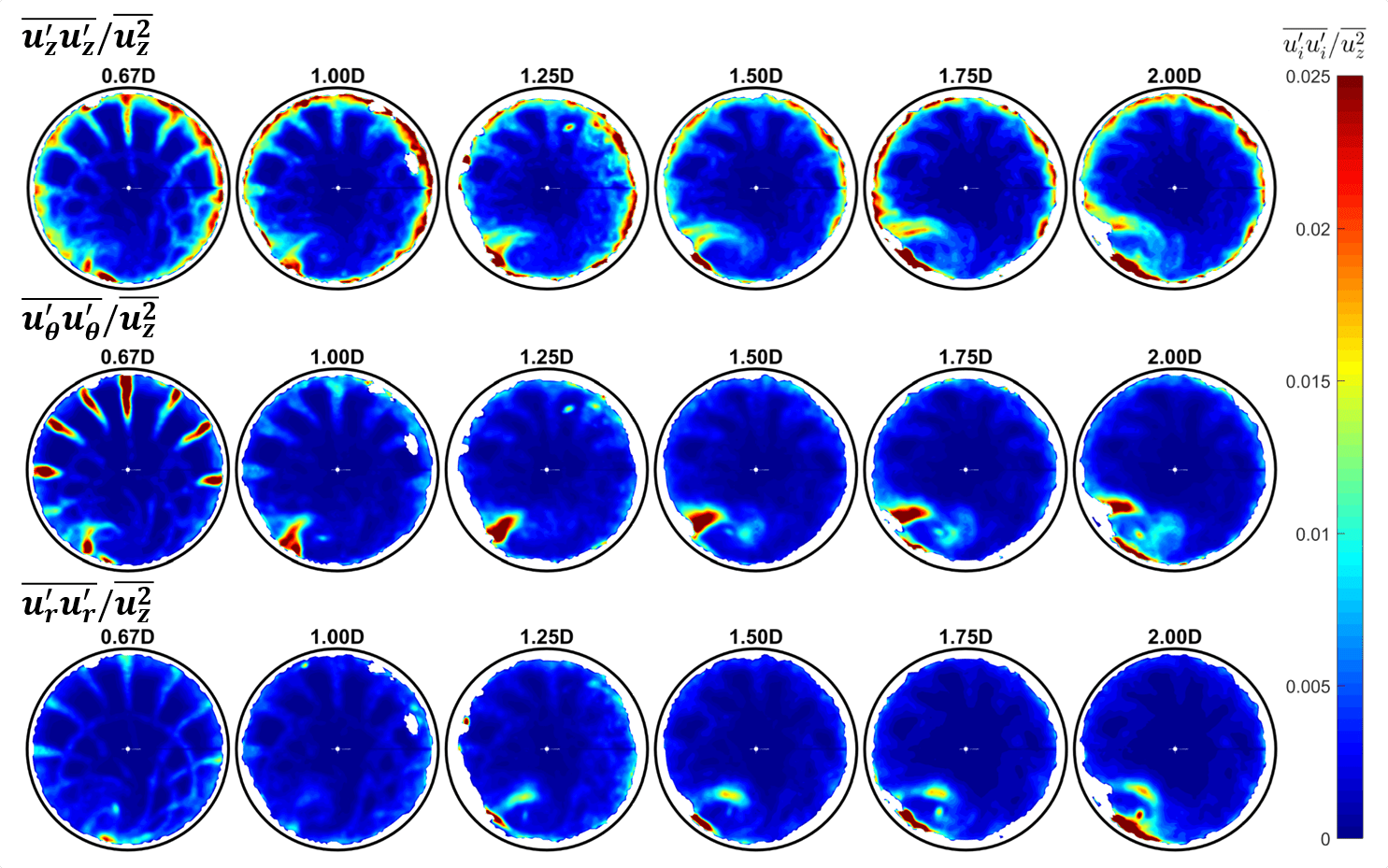}
\caption{Normal components of the Reynolds stress tensor in the axial ($\overline{u_{z}^{'}u_{z}^{'}}$) tangential ($\overline{u_{\theta}^{'}u_{\theta}^{'}}$) and radial ($\overline{u_{r}^{'}u_{r}^{'}}$) directions for the small-scale experiment, normalized by the average bulk axial velocity, $\overline{u_{z}^{2}}$}
\label{fig:14}
\end{figure}

Though the behavior of the single-vortex distorted flow is very similar across scales, there are details that are observed in the mean flow results of the full-scale engine test that were not immediately clear from the small-scale results, most likely due to the resolution of the PIV measurement system. As mentioned when analyzing the secondary flow results, at the 0.50D full-scale measurement plane, it is possible to observe indications of three distinct vortices formed from the StreamVane, which later combine into one stronger vortex. This is even more obvious when looking at visualizations present in the particle images; one such example is presented in Fig.\ref{fig:12}, and in the streamwise vorticity profile for the full-scale experiments. These tip vortices are generated by the overall shape of the vortex-generating feature in the StreamVane device shown previously in Fig. \ref{fig:2}, which consists of a ring that connects the tips of four longer turning vanes, similarly to what is observed in tip vortices generated by slotted bird wings \cite{KleinHeerenbrink2017}. This spreads the vorticity generated by the vortex, leaving room in the core of the vortex for the high axial velocity observed previously in Fig. \ref{fig:11}. Given that the regions outside of the inner core of the vortex are generated as expected, it is suggested that those tip tabs are either shortened or extended until meeting at the center of the ring for future StreamVane designs to prevent the generation of those tip vortices.\

\begin{figure}
\centering
  \includegraphics[width=0.6\textwidth]{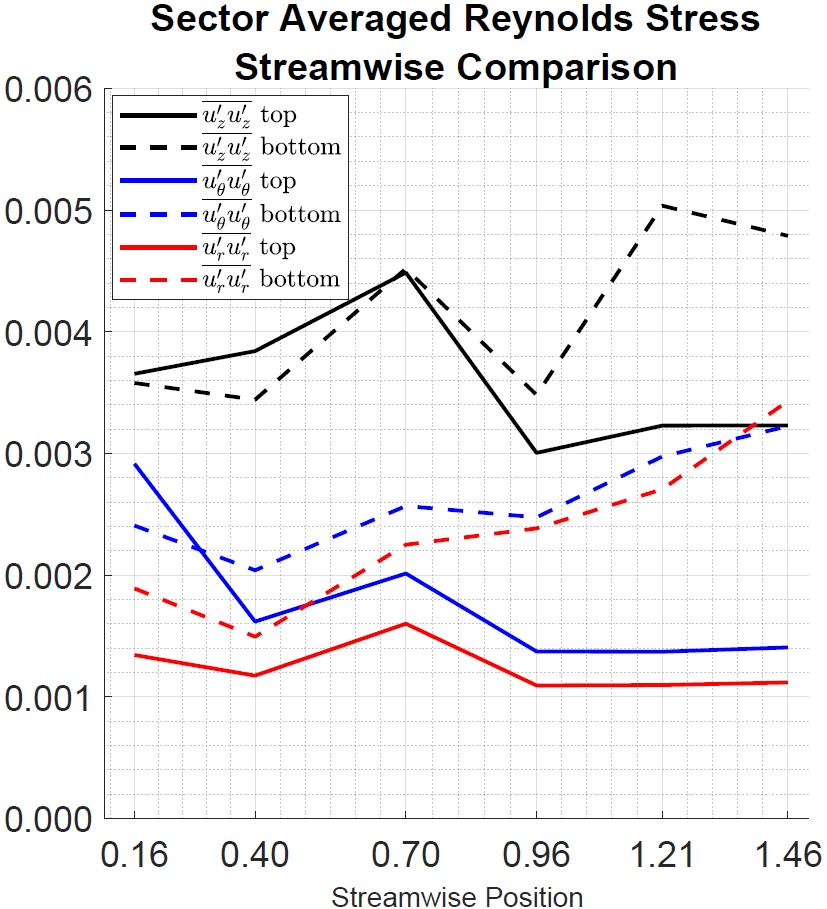}
\caption{Sector averaged Reynolds stress components. Results show averaged values for top and bottom halves of plots presented in Fig. \ref{fig:14}.}
\label{fig:15}
\end{figure}

Even though the discrete vortex structure was not obvious in the mean velocity profiles for small-scale, an analysis of the streamwise vorticity profiles of the small-scale experiment confirms that this also happens in that scale, as shown in Fig. \ref{fig:13}. At the 0.67D plane, it is possible to see three relatively distinct regions of higher vorticity, indicating those tip vortices before combining into one single stronger vortex. There is also a region of increasing vorticity following the vortex development, similar to that observed by Murphy and MacManus \cite{Murphy2011}, which is believed to originate from vorticity tilting from the boundary layer in the inside of the duct into axial vorticity, as described by Klemp \cite{Klemp1987}, providing a continuous convection of vorticity into the vortex core. \

To analyze the turbulent characteristics of the development of the vortex, it is necessary to first assess the instabilities related to its generation, as suggested by Devenport et al. \cite{Devenport1996}, and Murphy and MacManus \cite{Murphy2011}. A qualitative observation of the instantaneous snapshots (not shown) of the small-scale data at several planes, and the full-scale data at the 1.40D plane show that the vortex is stable as it is generated, suggesting that instabilities presented downstream of the generation are related to the turbulence intrinsic to the vortex development. To illustrate this, the normal components of the Reynolds stress tensor are presented in Fig. \ref{fig:14}. The regions of growing small-scale instabilities indicate the development of the vortex and growing turbulence associated with its transition to turbulence. Interestingly, as flow approaches the fan in a full-scale engine, the acceleration generated by the fan and core suction induce axial stretching of the vortex, increasing its stability and reducing its turbulent characteristics, as shown experimentally by Guimar\~{a}es et al. \cite{Guimaraes2018b}, confirming the numerical observations of Delbende et al. \cite{Delbende2002}, and Nolan \cite{Nolan2001}.\

One last point to make is regarding the appearance of local instabilities and turbulence due to the presence of vanes, even in regions further away from the vortex. The turbulence is higher at the plane immediately downstream of the StreamVane due to the intense local shear from the vane wakes, but these features rapidly mix and attenuate as flow develops downstream. By calculating the sector averaged stresses for each measurement location, shown in Figure \ref{fig:15}, it is possible to isolate the top region of the plots shown in Figure \ref{fig:14}, where the vane wakes are more pronounced (full lines), and the bottom region, where the vortex turbulence is more significant (dashed lines). Overall, there is a decrease in the average Reynolds stress level of the top half, and an increase in the level of the bottom half, supporting the claims presented before. For more details on the effects of small-scale StreamVane turbulence, refer to Guimar\~{a}es et al. \cite{Guimaraes2017b}, and for full-scale StreamVane turbulence refer to Guimar\~{a}es et al. \cite{Guimaraes2018a}. As concluded in these two aforementioned references, the turbulent length scales introduced to the flow by the StreamVane vanes are much smaller than the length scales of the vortical features generated by the StreamVane, thus providing additional representative realism to the secondary flow features generated. \

\bigskip
\section*{Conclusions}
\label{conc}
A single-vortex distortion was generated in two different scales: a small-scale low speed wind tunnel, and in the inlet of a full-scale engine testing rig. Three-component velocity data was taken with stereoscopic image velocimetry at pre-defined positions downstream of the generation of the distortion to analyze how the flow develops along a duct.\

Mean flow results of velocities and flow angles indicate that the generation and development of the vortex is similar across scales, not depending on the Reynolds number of the flow, and being governed mostly by two-dimensional vortex dynamics. Details of the StreamVane are responsible for an overshoot in the generation of the full-scale tangential angles, and swirl intensity in the flow, but the overall characteristics of the flow are generated well in both scales.\

A deeper analysis of the development of the vortex was done in relation to turbulence and vorticity in the small-scale results. Three tip vortices are generated right downstream of the StreamVane, and later combine into one single vortex core. With the development of the vortex, the instabilities in the flow around it increase, as expected for a vortex in an axial flow. Instabilities introduced to the flow by the StreamVane vanes are evident right downstream of the screen, but decrease as flow develops. \

The results of this work indicate that investigations of complex inflow distortions can be performed in small-scale wind tunnel environments, decreasing costs associated with full-scale engine tests. To improve these capabilities, it is recommended that small-scale wind tunnels are equipped with features simulating those of full-scale engines, such as spinners, rotors, and fans. 

\bigskip

\section*{Acknowledgements}
The authors would like to acknowledge NIA and NASA Langley Research Center for funding this work in association with NASA's Environmentally Responsible Aviation Project (NIA cooperative agreement RD-2917), project managers Fay Collier (LaRC), Hamilton Fernandez (LaRC), Greg Gatlin (LaRC), and Bo Walkley (NIA). An additional thanks to CAPES for financial support to Tamara Guimar\~{a}es.

\bigskip

\section*{Appendix: Errors in flow angle comparisons}
\label{App}
\begin{figure}[h!]
\centering
\includegraphics[width=0.32\textwidth]{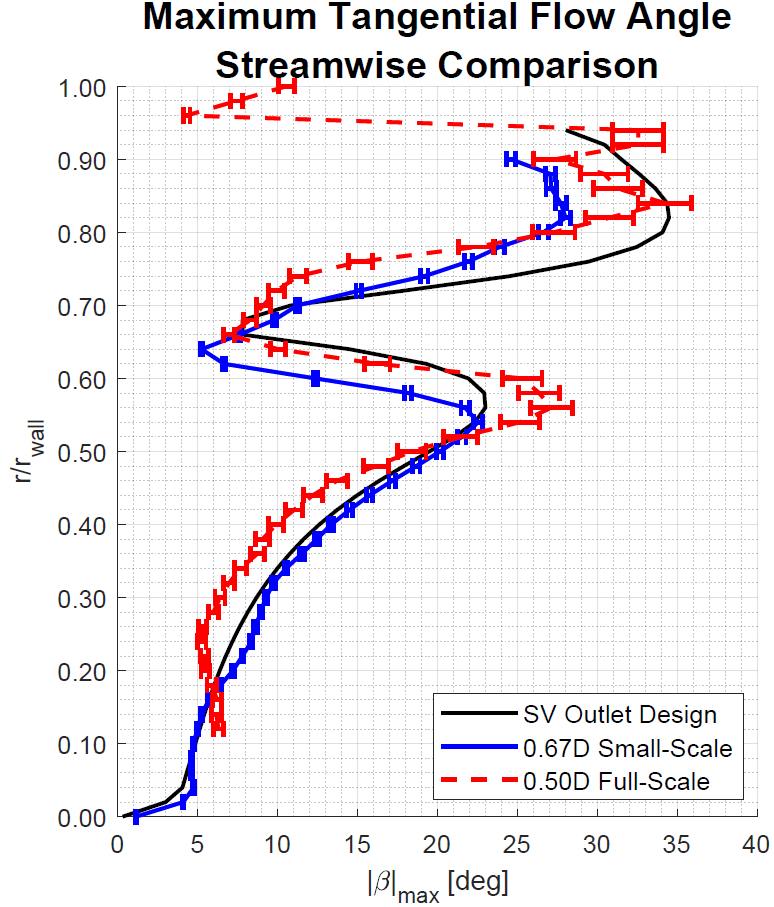}
\includegraphics[width=0.32\textwidth]{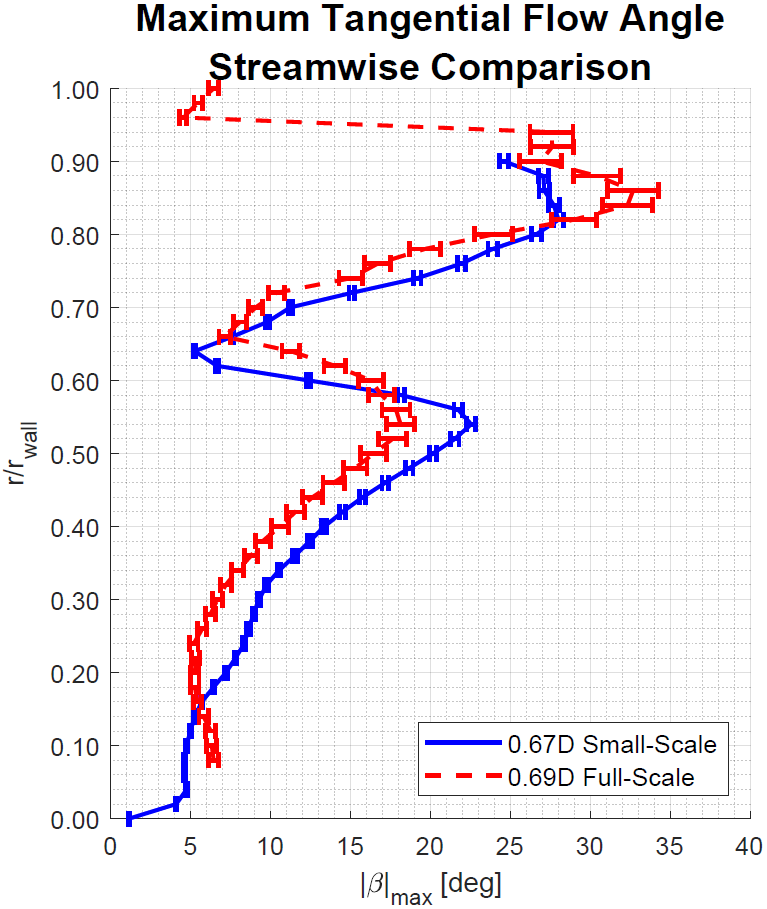}
\includegraphics[width=0.32\textwidth]{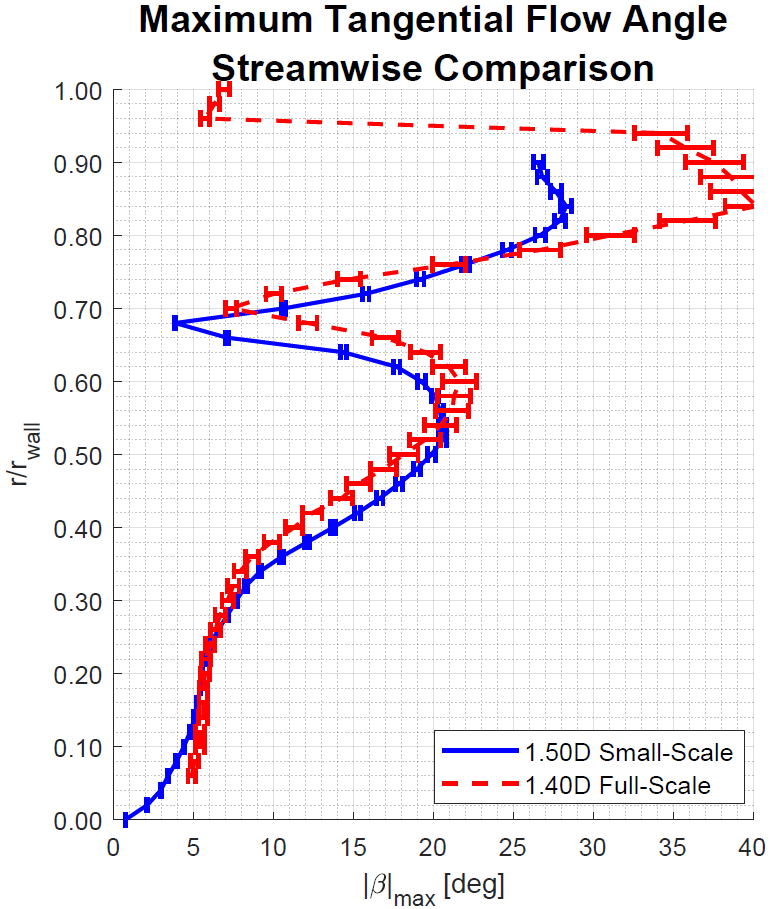}
\caption{Maximum tangential flow angle comparison between small and full-scale experiments at each plane with error bars.}
\label{fig:16}
\end{figure}

\begin{figure}
\centering
\includegraphics[width=0.32\textwidth]{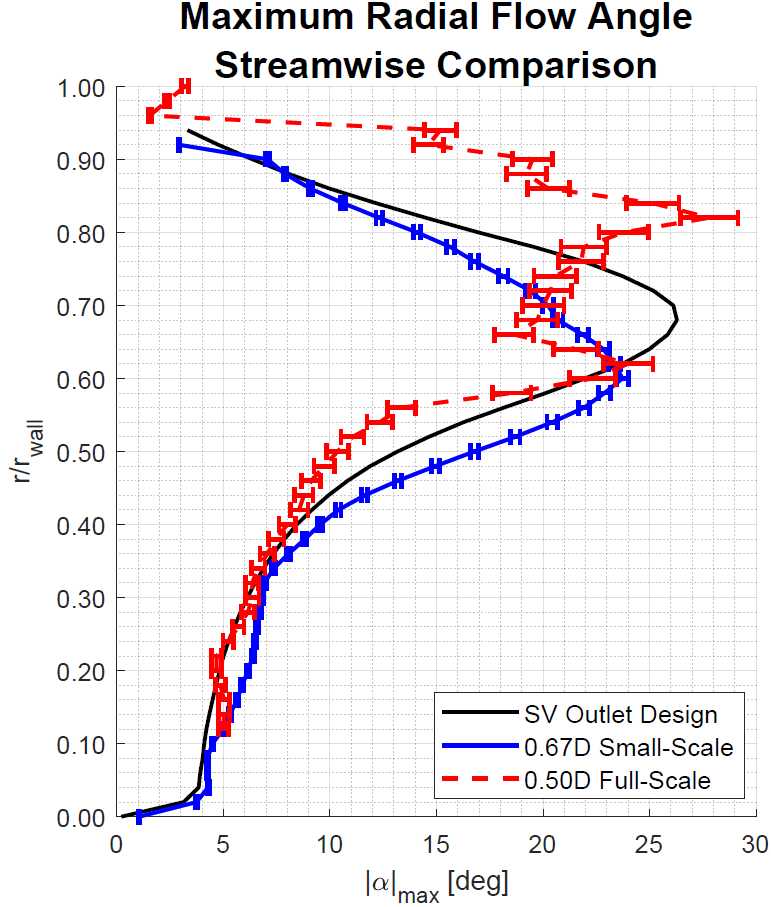}
\includegraphics[width=0.32\textwidth]{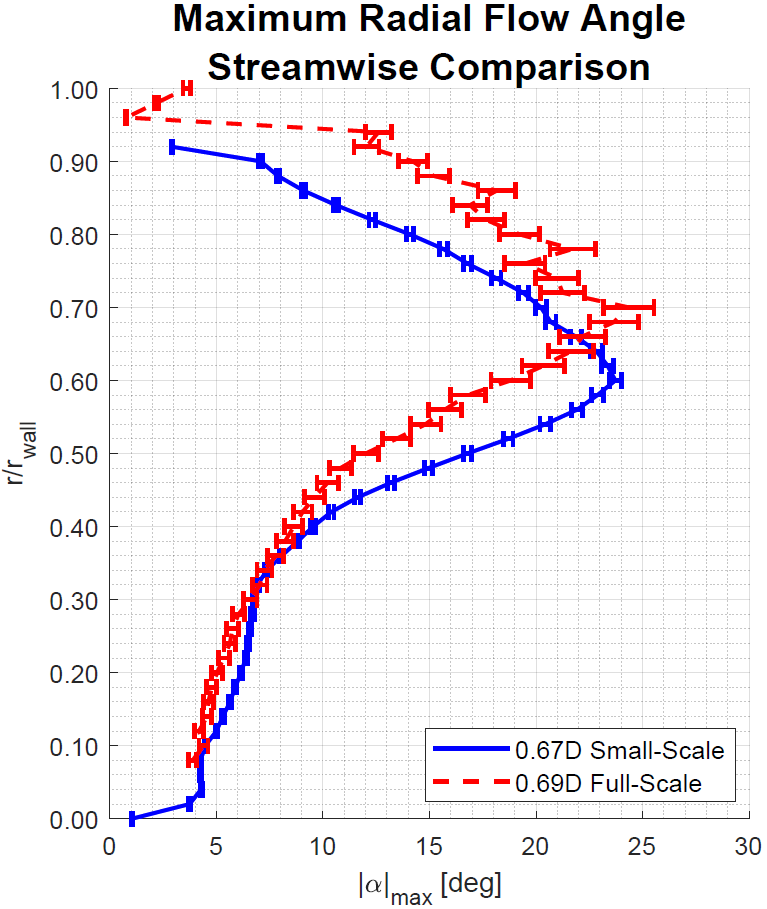}
\includegraphics[width=0.32\textwidth]{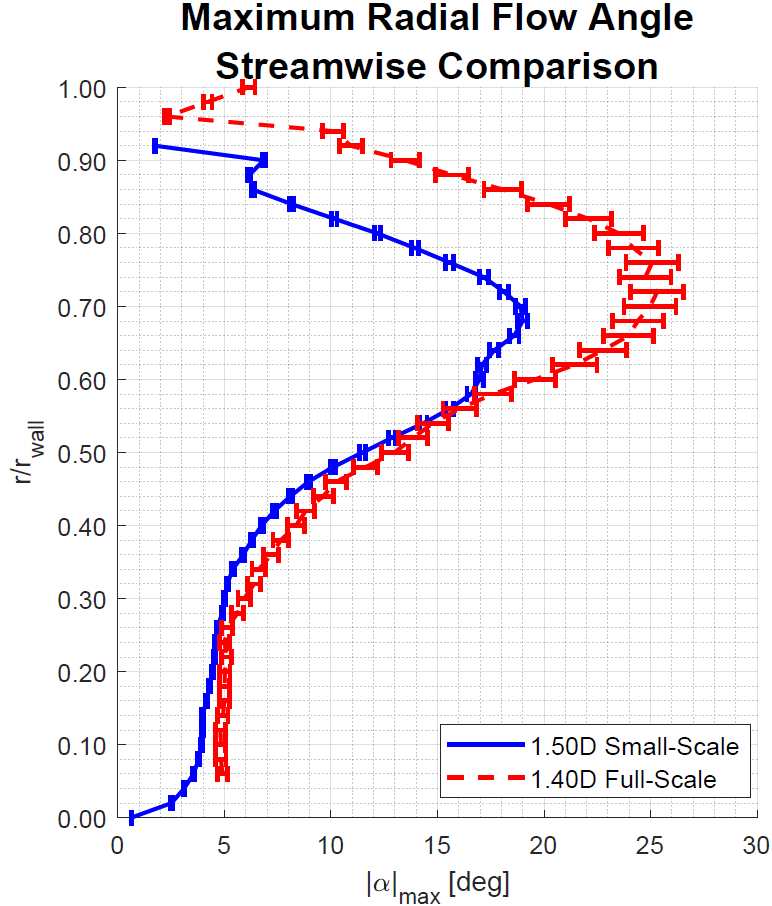}
\caption{Maximum radial flow angle comparison between small and full-scale experiments at each plane with error bars.}
\label{fig:17}
\end{figure}

This appendix (Figures \ref{fig:16} and \ref{fig:17}) presents the maximum flow angle comparisons between small and full-scale experiments when the error bars are added to the previously presented plots. The error is calculated from the uncertainty values presented previously ($\pm1.2$\% for small-scale, $\pm4.9$\% for full-scale).

\newcommand{\noopsort}[1]{} \newcommand{\printfirst}[2]{#1}
  \newcommand{\singleletter}[1]{#1} \newcommand{\switchargs}[2]{#2#1}

\end{document}